# The Right Tool for the Job

Paul Kienzle

*NIST Center for Neutron Research*
*Gaithersburg, MD  20899*

Tcl/Tk provides for fast and flexible interface design but slow and cumbersome vector processing.  Octave provides fast and flexible vector processing but slow and cumbersome interface design.  Calling Octave from Tcl gives you the flexibility to do a broad range of fast numerical manipulations as part of an embedded GUI.  We present a way to communicate between them.

# Tcl/Tk

User interface programming is hard no matter what language you are using.  Tcl/Tk makes it easier than most.

- **Rapid development**
  Can enter commands directly into a running application.   Can reconfigure widgets and replace procedures on the fly.
- **Easy to learn**
  Structured as a few commands with reasonable defaults and many options giving both ease of use and flexibility.
- **Extensive GUI support**
  Tk provides native look and feel.  Large base of contributed widgets available on the net.
  BLT provides a vector data structure and a very complete graph widget.
  Canvas widget allows you create sophisticated widgets in a Tcl script (like the file selection dialog shown here).
- **Excellent string handling**
  Powerful regular expression search and replace facility.  Fast enough that we can process file headers for 1000 files in a few seconds while the user waits.  We code file processing in as easily maintained scripts rather than compiled C.
- **Extensible**
  Well defined mechanism for defining new commands, so if scripts are too slow we can easily rewrite parts in C.  We haven't had to do so yet.
- **Cross Platform**
  Windows, Macintosh and many varieties of Unix.  Some extension packages are not available on every platform.
- **Community**
  The Tcl community is large and active.  Communication via newsgroup (news:comp.lang.tcl) and Wiki (http://mini.net/tcl).
- **Open source**
  Problems can be resolved locally.  Programs can be distributed without prejudice.

- **Flexible integration**
  Easy interprocess communication. Can write an echo client *and* server in a dozen lines of code. Can use Expect package to interface with existing command driven programs.
- **No numerical support**
  Even if we were to extend Tcl/Tk with numerical commands, it would pale in comparison to what is available already in Octave/Matlab.

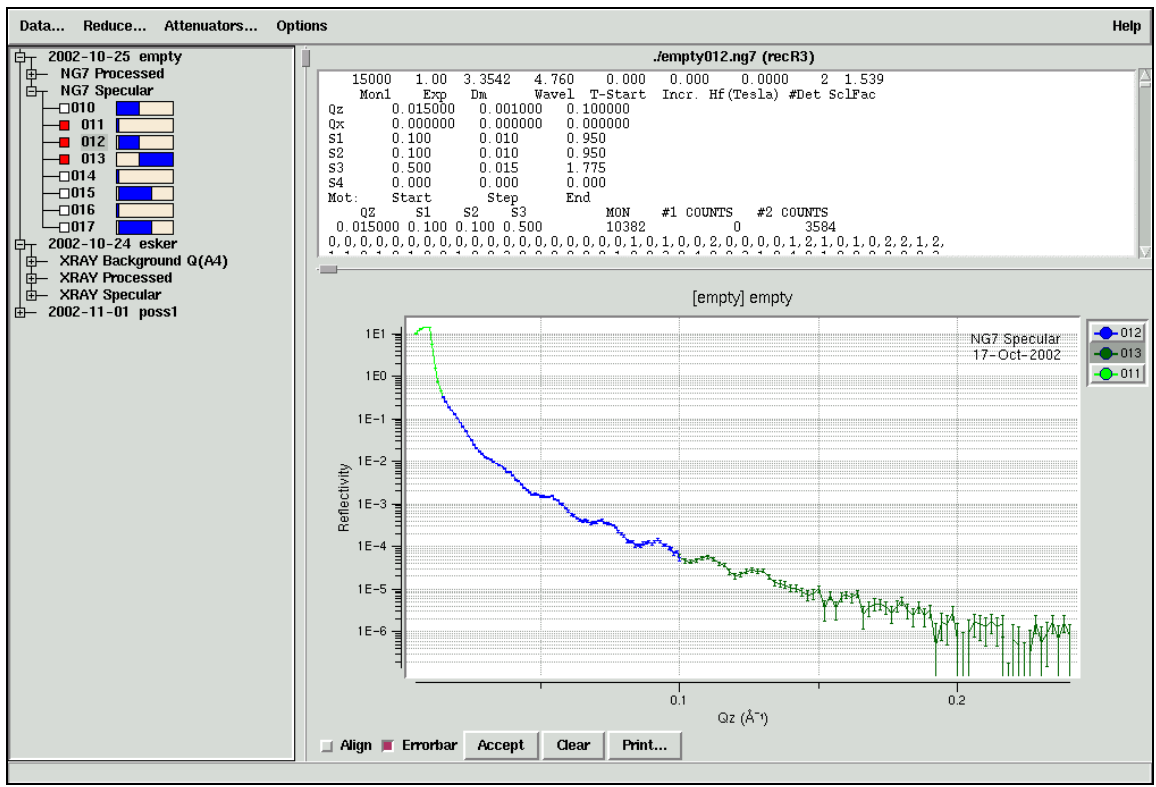

*Fig. 1. Tree selection widget from the BWidgets toolkit is implemented as a Tcl script using the Tk canvas widget. After loading, we replace a key function to provide a hook that lets us draw the Q range indicator bar beside the run number.*

# Octave

Creating fast, accurate and stable numerical algorithms is hard. Octave provides an easy interactive interface to LAPACK and other netlib software so that you don't have to.

- **Rapid development**
  Can enter commands interactively. Can replace functions in a running application.
- **Easy to learn**
  Simple procedural structure. Convenient data structures.
- **Extensive numerical libraries**
  Fourier transforms, interpolation, ODE's, integration, optimization, signal and image processing, control systems. Mostly compatible with Matlab so lots of code on the net can be made to run without difficulty.

- **Excellent matrix handling**
  Array slicing and reordering is fast and flexible. Convenient vector arithmetic. Sparse matrix solvers.
- **Extensible**
  Can interface to existing FORTRAN and C applications. Partial support for Matlab mex source.
- **Cross Platform**
  Support for Windows, Macintosh OS/X and Unix.
- **Community**
  Octave is under active development. Communication is via mailing lists (http://www.octave.org). Most questions are answered within hours.
- **Open source**
  Problems can be resolved locally. Programs can be distributed without prejudice.
- **No GUI toolkit**
  Even if we were to extend octave with a GUI toolkit, it would pale in comparison to what is available from Tcl/Tk.

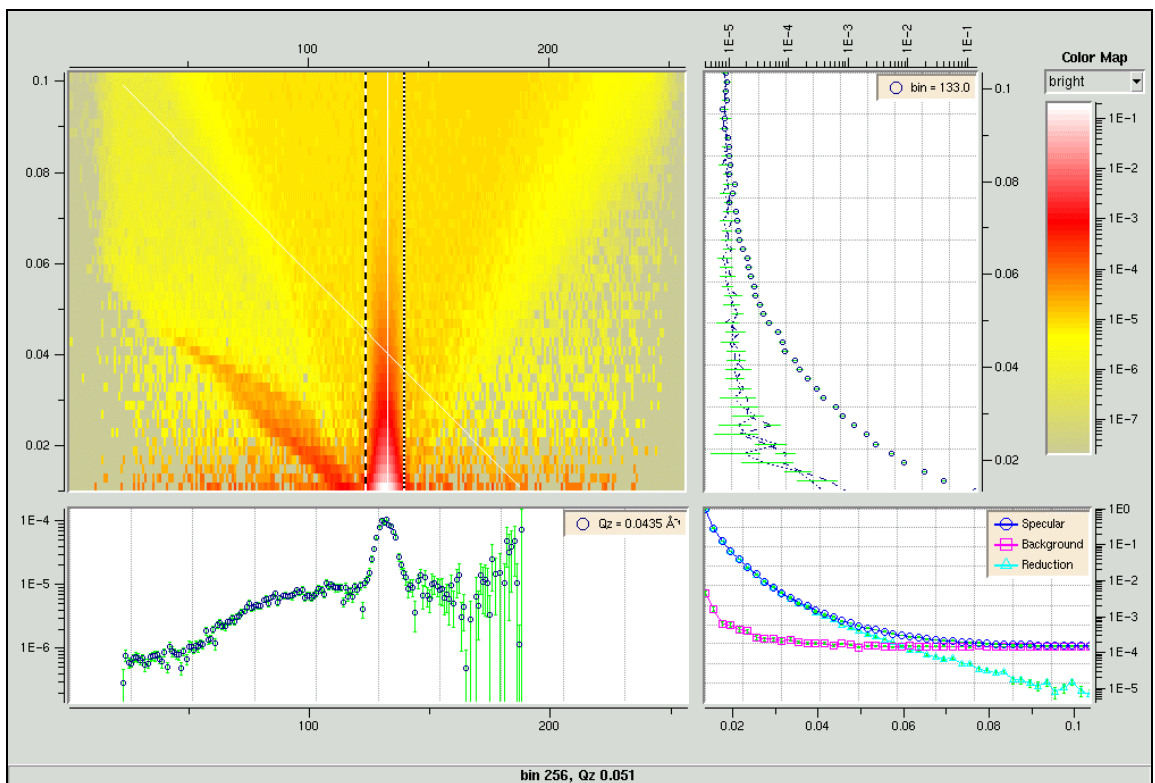

*Fig 2. Octave converts the matrix into a color image, extracts the cross section and integrates the specular region between the two dashed lines. BLT plots the graphs. Tcl/Tk handles layout and resizing. About 650 lines of code, including comments.*

# Tcl/Tk → Octave

Tcl/Tk and Octave are complementary. All we need is to communicate between them.

Tcl needs to send BLT vectors and strings, and receive BLT vectors, photo images and strings from octave.

- **octave connect**
  Start the communication.
- **octave send** *v x*
  Send the tcl variable *v* to the octave variable *x*.
- **octave recv** *v expr*
  Ask octave to send the value of the expression *expr* from octave into the Tcl variable *v*.
- **octave eval {** *commands* **}**
  Send commands to octave to process at its leisure.
- **octave capture {** *commands* **}**
  Send commands to octave and capture the output.  *** Not implemented ***
- **octave sync ?***callback* **?***args*
  Communication is assumed to be asynchronous.  The sync command without a callback assures that all **eval**'s and **recv**'s issued until now have been processed.
  If octave is being called in response to a callback, you need to be sure that only one request is being processed, and that all intermediate events are ignored until Octave is ready to process the next request. Sync with callback does that.
- **octave cancel**
  Cancel the current and all pending octave requests.  *** Not implemented ***
- **octave close**
  End the communication.

# Octave →Tcl/Tk

We wrote the octave end of the link as a generic server knowing nothing about the enviroment connecting to it.  In principle it could be another environment entirely such as Python or Java, or it could even be a client written in Octave.

- **listen (***port*, *hosts***)**
  Listen for connections from *hosts* on *port*.  Host-based authentication.
- **send ("***command***")**
  Send command string to client.  Octave does not know what is processing the commands at the other end.  There is no assumption that it is Tcl/Tk.
- **send("***name***",***expr***)**
  Send *expr* to *name*.  Client is expected to place the value into the named variable.  Octave indicates the type of the expression (currently string or matrix).

# Issues

- **Security**
  Making a sandbox to secure octave would be a significant undertaking, so we must assume trusted users only.  Host-based authentication is weak.  We should instead be using SSL or ssh to establish authenticity.
- **Process Control**
  There is at present no mechanism to interrupt long running Octave callbacks.  Because

Octave is potentially running on a different machine, we will need to set up a kill server which can accept interrupt requests when Octave is busy.
- **Speed**
Communication overhead and server response time are of concern. Because the responses are processed asynchronously from the event queue, we also have to wait for other events in the Tcl event loop to be processed, possibly causing flicker.
- **Maintenance**
Supporting this package requires someone familiar with both Octave and Tcl, however both languages are common and simple enough to learn.
- **Impedance mismatch**
There is potential for confusion between Octave and Tcl syntax, particularly in the use of quotes, brackets and braces. We can work around this with the appropriate escape sequences, but it will make the octave code in Tcl a little bit ugly. We have not yet tried to make Tcl easy to call from Octave.
- **Name space separation**
All Octave commands are evaluated in the top level namespace and all responses are evaluated in the Tcl top level namespace. Until we have better control of scoping we will need to be careful with the variable names that we use. Automatic sharing of variables would be convenient.

# Example

The following example shows how to normalize an uncorrected reflectivity signal. It is not atypical of the octave code you would find in our Tcl scripts.

```
octave eval {
    # There is a initial intensity drop-off in uncorrected data
    # followed by a gradual rise.  Skip over it when normalizing.
    up = find(diff(refl.y)>0);
    if !isempty(up) [peak,idx] = max(refl.y(up(1):length(refl.y)));
    else [peak,idx] = max(refl.y); endif
    dpeak = refl.dy(up(1)+idx-1);
    send (sprintf('set ::transmission_coeff %g',peak));
    send (sprintf('set ::dtransmission_coeff %g',dpeak));
    refl = run_scale(refl, 1/peak, dpeak / peak^2);
    send_run("refl_%s", refl);
}
```

# Conclusions

Tcl/Tk profides fast and flexible interface design. Octave provides fast and flexible numerical processing. While not without its warts, using Octave as a compute engine from Tcl allows us to create sophisticated scientific applications with minimal effort.